\documentclass[rnote,traditabstract]{aa}
\usepackage{natbib}
\usepackage{graphicx}
\bibpunct{(}{)}{;}{a}{}{,}

\newcommand{\sz}{\hbox{SZ~Cam }}

\newcommand{\ks}{km~s$^{-1}$}
\newcommand{\vsin}{$V_{\rm rot}\sin i$ }

\newcommand{\ms}{M$_{\odot}$}
\newcommand{\rs}{R$_{\odot}$}

\newcommand{\he}{He\,{\sc i}\,\,6678 }
\newcommand{\hea}{He\,{\sc i}\,\,6678}

\newcommand{\Aa}{\accent'27A}

\begin{document}

   \title{The O-type eclipsing binary \sz revisited
\thanks{Based on spectral observations from
the Ond\v rejov Observatory and on observations collected at the German-Spanish
Astronomical Centre, Calar Alto, operated by the Max-Planck-Institute f\"ur Astronomy, 
Heidelberg, jointly with the Instituto de Astrofisica de Andalusia.}
}

\author{Pavel~Mayer\inst{1}
   \and Horst~Drechsel\inst{2}
   \and Ji\v r{\'\i{}}~Kub\'at\inst{3}
   \and Miroslav \v Slechta\inst{3} }

   \offprints{P. Mayer, \email{mayer@cesnet.cz}}

  \institute{
   Astronomical Institute of the Charles University,
   Faculty of Mathematics and Physics, V~Hole\v{s}ovi\v{c}k\'ach~2,
   180 00 Praha~8, Czech Republic
\and       Dr. Remeis-Sternwarte, Astronomisches Institut der Universit\"at
	   Erlangen-N\"urnberg, Sternwartstra\ss e 7, D-96049
	   Bamberg, Germany
\and       Astronomical Institute, Academy of Sciences of the Czech Republic,
           Fri\v cova 298, 251 65 Ond\v rejov, Czech Republic           }

\date{Received \today  / Accepted}

 \abstract{We analyse new spectra of the multiple system \sz because
previous studies found different values of the 
primary radial velocity amplitude. The older solutions of light 
curves also have different ratios of secondary to primary luminosity as
inferred from the observed equivalent widths of spectral lines. We
therefore reanalyse the light curves of the eclipsing pair. Only the
light curve derived by Wesselink has a solution that agrees with the
observed equivalent width ratio. The resulting parameters of the binary are
discussed. Masses of
$M_1=16.6$ and $M_2=11.9$ M$_{\odot}$, and radii $R_1=9.4$ and 
$R_2=5.4$ R$_{\odot}$ are derived. We point out that radial
velocities measured with the CCF method can be misleading when the
method is applied to multiple systems with complex line blends. New 
radial velocities are also obtained for the visual component ADS~2984~A
(HD~25639).

}
   \titlerunning{The O-type eclipsing binary \sz}
   \authorrunning{P. Mayer et al.}

   \maketitle
%
%________________________________________________________________

\section{Introduction}
The early-type eclipsing binary SZ~Cam (HD~25638) has been the subject
of several studies in recent years. The binary is a member of the open
cluster NGC~1502 and is also the northern component of the visual binary
ADS~2984. With a combined spectral type of O\,9.5 and 
detached components, the binary is a potentially important source of
absolute parameters of massive stars. However, the star was found to be
a triple system -- or, more specifically, a system consisting of two
close binaries \citep{m94}. The third contributor to the light makes the
light curve solution difficult, and also causes problems in the analysis
of the complex spectra.

The third body in \sz manifests itself in various ways: as a 
third component in the spectral line profiles, by the presence of the
light time effect, and by additive light in the light curves 
\citep[][hereafter LMD]{LMD}. The third body was also found independently
by speckle interferometry \citep{mason}. A new speckle measurement
was published by \citet{g2007}, and the magnitude difference was
given by \citet{balega} as $0\fm95\pm0\fm02$ at $\lambda 535$ nm.
In the Washington Double Star Catalog\footnote
{http://ad.navy.usno.mil/wds}, \sz is referred to as
component Ea, and the speckle component as Eb.

In addition to these studies, the star was analysed by \citet[][hereafter
HHH]{HHH}, \citet{g2000,g2002,g2008}, and \citet{mich}. HHH found
some basic parameters of the system, namely the semiamplitude $K_1$,
which differs from those published by LMD. 

We collected new spectra with the 2\,m telescope at the Ond\v rejov 
Observatory and also derived new solutions of the light curves.
Based on these data, we expect to resolve the discrepancy between HHH and
LMD. The already published results and our new results are listed in 
Table~\ref{RES}. Note that the radial velocity (RV) semiamplitudes $K_1$
found by \citet{mich} and by \citet{g2008} are close to those of LMD.
In this paper, the linear ephemeris given by \citet{g2007} is
used:

Primary minimum = HJD $2453676.1628 + 2\fd6984222$.\\

\begin{table*}
\caption{The published parameters of the RV curve (values in \ks).}
\label{RES}
\begin{tabular}{lllllrll}
\hline\noalign{\smallskip}
Source                & $K_1$ & m.e. & $K_2$ & m.e. & $V_{\gamma}$ & m.e. & Method                     \\[1mm]
\hline\noalign{\smallskip}
Harries et al. 1998   & 225.8 & 3.8  & 259.1 & 4.0  &$-12.0$       & 3.0  & CCF method                 \\
Lorenz et al. 1998    & 180.2 & 2.0  & 261.2 & 3.8  &$ -2.9$       & 1.6  & MIDAS Gaussian fitting     \\
Michalska et al. 2007 & 181.6 &      & 268.2 &      &              &      & KOREL applied to H$\alpha$ \\
Gorda 2008            & 192.0 & 2.6  & 266.4 & 2.5  &$-10.6$       & 2.0  & Gaussian fitting           \\
This paper            & 189.4 & 1.4  & 264.1 & 2.5  &$ -2.3$       &      & Gaussian fitting           \\
\hline\noalign{\smallskip}
\end{tabular}
\end{table*}

\begin{figure}
\includegraphics[width=8cm]{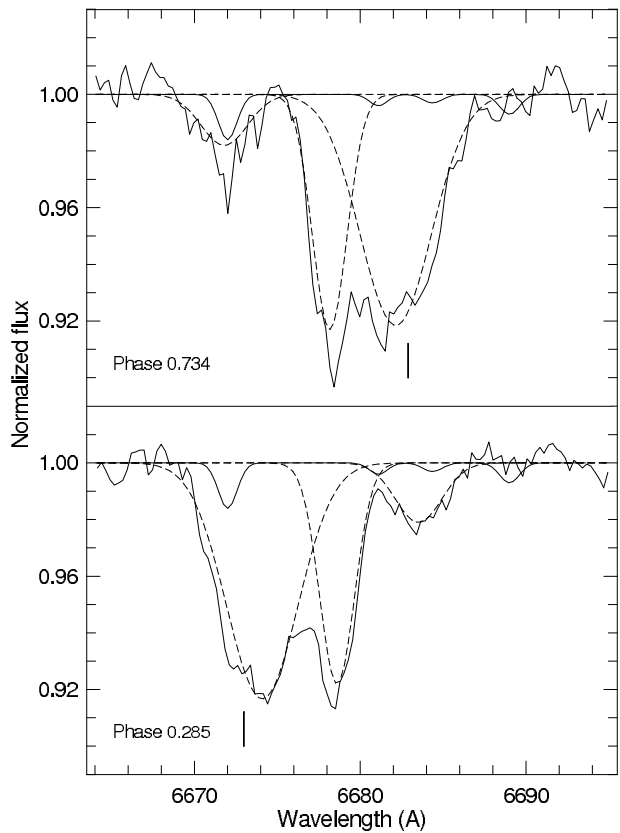}
\caption{Profiles of the line \hea~\Aa\, taken close to opposite 
 quadratures, at phases 0.285 and 0.734. The dashed lines represent
 the Gaussian fits of all three components of the system (the wide 
 profile -- primary, the shallow one -- secondary, the narrow one -- 
 tertiary). The short vertical lines show the expected primary line 
 positions according to the HHH solution. Four superimposed DIBs are
 also indicated.}
\label{EXAMPLES}
\end{figure}

\section{Spectroscopy}
% \subsection{New radial velocities}
The spectra analysed by LMD were obtained with the 2.2\,m telescope at
the Calar Alto Astronomical Centre (hereafter CA). New spectra for \sz
were recorded with the coud\'{e} spectrograph at the 2\,m telescope of
the Ond\v rejov Observatory. The region between 627~nm and 678~nm is
covered by 46 spectra with a two-pixel resolution of 12700 and a typical
signal-to-noise ratio (S/N) of
$\approx 200$. There are also two echelle spectra covering the region 
from 400~nm to 670~nm publicly available at the ELODIE archive 
\citep{moul}. The spectral region of the \hea~\Aa\, line in two
Ondrejov spectra taken at opposite quadrature phases is shown in 
Fig.~\ref{EXAMPLES}. The relatively narrow and strong third-body line is
clearly visible in the centre of the blend. Owing to the
strong reddening ($E_{B-V}=0.68$), several DIBs are present: a stronger
one at 6672.0~\Aa\, with equivalent width (EW) of about 0.02~\Aa\,
and several other weaker ones at 6681.1, 6684.4, and 6689.0~\Aa\, (the
wavelengths of these DIBs were measured in our spectra).\footnote{At 
conjunctions, the DIBs 6672.0~\Aa\, and 6689.0~\Aa\, are not blended
with the binary lines and can be measured relatively well. The other
two DIBs are always superimposed on stellar profiles; their parameters
were determined by optimizing the multiple Gaussian fit of the
composite profile.} Two short vertical lines mark the primary radial 
velocities (RVs) as they follow from the RV curve solution by HHH. 
These RVs are clearly incompatible with our
spectra, and the large value of $K_1$ published by HHH is not supported.

We fitted the line profiles by Gaussians. The resulting RVs are listed 
in Table~2. Owing to the complicated structure of the profile, reliable 
results can only be expected from spectra taken near quadratures, when
the secondary line is separated. At phases around 0.75, the secondary
component is blended with the DIB at 6672~\Aa. Close to the He\,{\sc i} 
line, there is the He\,{\sc ii} line 6683~\Aa. We did not detect any
contribution of this line, although in other stars of nearly identical
spectral type -- but probably of higher luminosity -- the He\,{\sc ii} 
line plays a role (e.g. at $\delta$~Ori).

\begin{table}
\caption{New radial velocities.}
%\begin{minipage}[]{\textwidth}
\label{GAUSS}
\begin{tabular}{llrrlr}
\hline\noalign{\smallskip}
 HJD                     & Phase      & RV     & RV     & Phase   & RV    \\
-2400000                 &  ecl.      & pri.   & sec.   & third   & third \\[1mm]
\hline\noalign{\smallskip}
 53706.2888              & 0.1625     &$ -166.2$&  216.2 &  0.0757 &$-18.5$\\
 53706.3075              & 0.1694     &$ -164.9$&  222.1 &  0.0824 &$-26.7$\\
 53706.3777              & 0.1954     &$ -175.9$&  249.7 &  0.1074 &$-21.2$\\
 53706.4290              & 0.2145     &$ -191.4$&  233.3 &  0.1259 &$-17.7$\\
 53713.1984              & 0.7231     &   198.4 &$-261.0$&  0.5460 &  34.0 \\
 53713.2673              & 0.7487     &   179.2 &$-261.8$&  0.5707 &  30.2 \\
 53743.2557              & 0.8619     &   152.3 &$-181.9$&  0.2924 &  14.8 \\
 53744.2390              & 0.2263     &$ -182.5$&  242.3 &  0.6439 &  21.4 \\
 53747.3102              & 0.3645     &$ -140.4$&  192.9 &  0.7419 &$ -1.9$\\
 53817.4299              & 0.3500     &$ -174.1$&  197.2 &  0.8118 &$-20.9$\\ 
 53818.4625              & 0.7325     &   179.2 &$-285.2$&  0.1810 &   0.5 \\
 53832.4657              & 0.9220     &   108.0 &$-131.7$&  0.1875 &$ -0.7$\\
 53938.4909              & 0.2135     &$ -182.6$&  242.7 &  0.0946 &$-19.7$\\
 53938.5082              & 0.2200     &$ -182.0$&  237.8 &  0.1007 &$-18.1$\\
 53938.5465              & 0.2341     &$ -182.4$&  250.4 &  0.1144 &$-15.4$\\
 53938.5810              & 0.2469     &$ -183.7$&  262.5 &  0.1267 &$-15.0$\\
 54014.3879              & 0.3400     &$ -163.3$&  200.5 &  0.2299 &$ -9.1$\\
 54014.4173              & 0.3510     &$ -170.4$&  203.2 &  0.2404 &$ -6.4$\\
 54020.5757              & 0.6331     &   152.5 &$-207.1$&  0.4421 &  25.7 \\
 54028.3257              & 0.5051     &$  -42.8$&   68.6 &  0.2130 &$-29.3$\\
 54028.3496              & 0.5140     &$  -44.1$&   68.5 &  0.2216 &$-25.6$\\
 54028.4000              & 0.5327     &$  -21.7$&   78.4 &  0.2396 &$-27.6$\\
 54085.4032              & 0.6573     &   163.9 &$-216.5$&  0.6199 &  23.7 \\
 54085.4771              & 0.6847     &   178.2 &$-241.6$&  0.6462 &  25.4 \\
 54085.5942              & 0.7281     &   185.7 &$-267.2$&  0.6881 &  15.6 \\
 54085.6617              & 0.7532     &   184.4 &$-286.1$&  0.7123 &  12.2 \\
 54085.7320              & 0.7792     &   192.0 &$-276.2$&  0.7374 &  12.2 \\
 54097.2111              & 0.0332     &$  -51.2$&   51.5 &  0.8416 &$-16.7$\\
 54097.2940              & 0.0638     &$  -94.7$&   87.4 &  0.8712 &$-17.2$\\
 54097.3726              & 0.0930     &$ -107.9$&  118.8 &  0.8993 &$-15.0$\\
 54097.4929              & 0.1376     &$ -155.8$&  213.1 &  0.9423 &$-28.5$\\
 54097.6107              & 0.1813     &$ -180.6$&  221.6 &  0.9844 &$-31.8$\\
 54097.7207              & 0.2220     &$ -184.4$&  253.0 &  0.0237 &$-26.8$\\
 54166.4347              & 0.6865     &   188.5 &$-286.5$&  0.5909 &  30.5 \\
 54166.4617              & 0.6965     &   188.8 &$-269.9$&  0.6006 &  29.9 \\
 54166.5121              & 0.7152     &   186.4 &$-277.1$&  0.6187 &  27.3 \\
 54166.5434              & 0.7268     &   189.1 &$-278.0$&  0.6298 &  27.3 \\
 54533.4269              & 0.6890     &   185.6 &$-240.3$&  0.8011 &   2.6 \\
 54533.4716              & 0.7056     &   185.9 &$-263.6$&  0.8171 &$ -3.3$\\
 54581.3352              & 0.4431     &$  -85.4$&  109.0 &  0.9297 &$-15.5$\\
 54590.3348              & 0.7785     &   188.2 &$-245.7$&  0.1473 &$ -2.0$\\
 54594.4061              & 0.2874     &$ -182.4$&  240.9 &  0.6029 &  20.4 \\
 54620.3787              & 0.9124     &   114.8 &$-142.4$&  0.8889 &$-21.0$\\
 54798.6220$^\mathrm{a}$ & 0.9668     &    35.8 &    2.6 &  0.6160 &  29.9 \\
 54932.4342              & 0.5562     &    63.2 &$ -94.4$&  0.4575 &  22.1 \\
 55060.5181              & 0.0222     &$  -55.2$&   57.3 &  0.2512 &   4.4 \\
\hline\noalign{\smallskip}
\end{tabular}
\begin{list}{}{}
\item[$^\mathrm{a}$] both SZ Cam and ADS~2984~A exposed in same frame
\end{list}
%\end{minipage}
\end{table}

\begin{figure}
\includegraphics[angle=270,width=8cm]{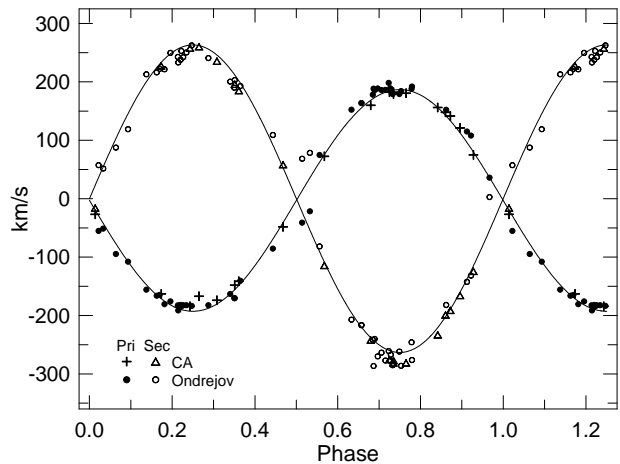}
\caption{Radial velocities of the eclipsing pair (the curves represent
 the solution with common $V_{\gamma}$).}
\label{ECLRV}
\end{figure}

We also refitted the CA spectra using an identical procedure. The RVs of
the eclipsing components are shown in Fig.~\ref{ECLRV}. The solution of 
the RV curves is listed in Table~\ref{SOLUTION}. Only RVs obtained using
the line \hea~\Aa\, from spectra taken in Ondrejov and Calar 
Alto were used; velocities from phases closer than 0.08 to both 
conjunctions were not considered. In the covered spectral range,
only the H$\alpha$ line is present in addition to the 6678~\Aa\, line. 
However, owing to its large width and disturbance by atmospheric lines
it is not suitable for Gaussian fitting.

In Table~2, RVs of the third lines are also contained. Since our
new solution of the third-body orbit does not differ considerably from 
that in LMD, we do not discuss these RVs here. Several spectra of the
component ADS 2984~A were also obtained, and the RVs are given in Table~4.
As the lines are wide, the resulting RVs are of low accuracy and the
period of this probable binary has not yet been able to be determined.

% \subsection{Spectral type, EW and FWHM}
Among the existing classifications of SZ~Cam, probably the most reliable
is the one according to \citet{budd}: O\,9.5\,V. This is the combined
spectral type. Following the scheme of \citet{conti}, we confirmed this
by comparing the lines 4471/4541~\Aa\ (spectral type) and 4089/4143~\Aa\
(luminosity) in the ELODIE spectra. However, a more detailed spectral
type classification of the secondary and tertiary components using
the ELODIE spectra is not possible due to weakness of the components'
spectral features. These components are probably main-sequence stars,
therefore their types and temperatures follow from their magnitude 
differences from the primary. Both should be later than the primary, so
the primary might be somewhat earlier than the type of the
integral spectrum. Its luminosity is also higher, since the radius is 
larger than the radius of a MS member as shown in Sect. 3; its
$\log g$ is 3.7, a value typical of the luminosity class IV \citep{mar}.
Although in stars of earlier spectral type the He\,{\sc ii} lines are 
usually observed, this is not so in this case because the strengths of
both the 6406 \Aa\, and 6683 \Aa\, 
lines strongly depend on $\log g$. According to \citet{lanzO},
the lines are absent for $\log g$ larger than 3.25 at $T_{\rm eff}=30500$~K.

\begin{table*}
\caption{Solution of the radial velocity curve of the eclipsing pair.}
\label{SOLUTION}
\begin{tabular}{lrrrr}
\hline\noalign{\smallskip}
&\multicolumn{2}{c}{Individual $V_{\gamma}$}&\multicolumn{2}{c}{Common $V_{\gamma}$}    \\[1mm]
\hline\noalign{\smallskip}
Parameter              & Primary         & Secondary       & Primary        &Secondary  \\[1mm]
\hline\noalign{\smallskip}
$K$ (\ks)              & 189.3$\pm 1.2$  & 263.4$\pm 1.8$  & 189.4$\pm 1.4$  & 264.1$\pm 2.5$  \\
$V_{\gamma}$ (\ks)     &   1.8$\pm 1.0$  &$-10.4\pm 1.6$   & $-2.3$\,\,\,\,\,\,& $-2.3$\,\,\,\,\,\,\\
Average $O-C$ (\ks)    &   5.1           &  11.5           &   7.0           &   12.6          \\
$A \cdot \sin i$ (\rs) &  10.09$\pm 0.09$&  14.04$\pm 0.14$&  10.10$\pm 0.10$&   14.08$\pm 0.1$\\ 
\hline\noalign{\smallskip}
\end{tabular}
\end{table*}

\begin{table}
\caption{Radial velocities of ADS 2984 A.}
\begin{minipage}[]{\textwidth}
\label{ADS}
\begin{tabular}{lrlr}
\hline\noalign{\smallskip}
HJD                     & RV    & HJD        & RV    \\
-2400000                & \ks   & -2400000   & \ks   \\[1mm]
\hline
53666.3657              &  23.2  & 54166.5829 & $-31.1$ \\
53670.3298              &  23.2  & 54170.4173 & $ -8.4$ \\
53680.6565$^\mathrm{a}$ &  10.0  & 54523.3472 &   20.1  \\ 
53797.4386              &  12.4  & 54593.3397 &   18.3  \\
53818.4075              &$-47.3$ & 54798.6220$^\mathrm{b}$ & $-17.2$ \\
53938.5452              &$-10.5$ & 55060.5444 &    1.7  \\
53947.4385              &  13.3  & 55074.4146 & $-51.3$ \\
54028.4483              &$-26.0$ & 55075.6340 & $-49.5$ \\
54166.4878              &$-15.6$ &            &         \\
\hline\noalign{\smallskip}
\end{tabular}
\begin{list}{}{}
\item[$^\mathrm{a}$] spectrum from the ELODIE archive
\item[$^\mathrm{b}$] both SZ Cam and ADS~2984~A exposed in same frame
\end{list}
\end{minipage}
\end{table}

\begin{table*}
\caption{Solution of various light curves.}
\begin{minipage}[]{\textwidth}
\label{LC}
\begin{tabular}{lccccc}
\hline\noalign{\smallskip}
Para-      & HHH\footnote{Solution by HHH; they used the two-filter data by \citet{cho}.}& Chochol&
LMD\footnote{LMD WD solution of Kitamura \& Yamasaki $UBV$ curves.} & Chochol   & Wesselink  \\
meter      &            & curves\footnote{New solution of Chochol's curves with WD code and
the mass-ratio $q$ used by HHH.}
  &           & curves\footnote{New solution of Chochol's curves with our $q$.}& \\[1mm]
\hline\noalign{\smallskip}
$L_3$\footnote{Assuming $L_1+L_2+L_3=1$.}
& 27.4\%\footnote{Calculated from absolute visual magnitudes in HHH Table 6.}   
                        & 24.9\%  & 31.0\%$\pm 1.5$ &  24.7\%$\pm 3.8 $& 30.9\%$\pm 1.8$    \\
$q$        &  0.871     &  0.871  &  0.704$\pm 0.015$ &   0.710$\pm 0.064$&   0.718         \\
$T_1$\,(K) & 29\,725    & 30\,500 & 30\,500   &   30\,500 &   30\,500                       \\
$T_2$\,(K) & 27\,183    & 27\,200 & 25\,370$\pm 230$\footnote{Recalculated for $T_1=30\,500$\,K.}
                                              &   27\,270$\pm 650$&  25\,300$\pm 740$       \\
$i$ (deg)  & 76.9$\pm 0.4$& 76.7  & 77.0$\pm 0.6$    &   77.0$\pm 1.8  $ &  79.7$\pm 1.3$   \\
$L_2/L_1$ in $V$ &$0.773^f$ &  0.536  &  0.422$\pm 0.031$&   0.492$\pm 0.054$& 0.262$\pm 0.029$ \\   
$r_1/A$    &  0.319$\pm 0.010$& 0.344& 0.380$\pm 0.030$& 0.353$\pm 0.016$& 0.379$\pm 0.004$ \\
$r_2/A$    &  0.297$\pm 0.012$& 0.270& 0.280$\pm 0.030$& 0.265$\pm 0.021$& 0.217$\pm 0.017$ \\
\hline\noalign{\smallskip}
\end{tabular}
\end{minipage}
\end{table*}

\section{New solutions of light curves}
\label{SLC}
In the following, we consider several \sz light curves:
\begin{itemize}
\item a photographic curve by \citet{wess}, the quality of which is
comparable to photoelectric curves, since it is based on more than
12000 measurements;
\item $UBV$ curves by \citet[][K-Y curves]{KY};
\item two curves obtained with intermediate wide filters centred on
4720 and 5720~\Aa\, by \citet{cho};
\item $UBVR$ curves by \citet{g2000} obtained with an area scanner, and
published in a graphical form only.
\end{itemize}
Owing to the close vicinity of ADS~2984~A and E, their {\it Hipparcos}
photometry \citep{ESA} is unhelpful.

These light curves differ for two possible reasons:
\begin{itemize}
\item the light curve is variable, or
\item the measurements are not reliable.
\end{itemize}
The variability of stars of the given parameters is not common. There 
are no indications that the magnitude in maximum changes as it would be
in the case of temperature or dimension changes. This explanation is 
therefore improbable.

Observing SZ Cam with a standard PMT photometer was not easy, because of
the close vicinity of the visual component. The K-Y and Chochol curves 
differ, each other and from the photographic curve. Gorda shows that
there were deviations from the average curve in some nights. However, 
more data would be needed to verify the changes; the variability
of the comparison star might be also the reason. The photographic curve
was collected over more than 6 years, which had the advantage
of smoothing the eventual short-time light curve variability.

A set of solutions is presented in Table~\ref{LC}. In our solutions 
$T_1=30500$ K was assumed. A value close to this was already used by
HHH, and according to newer temperature scales \citep[e.g.][]{massey}
it is a reasonable value for the spectral type of the primary component.

\begin{figure}
\includegraphics[width=8cm]{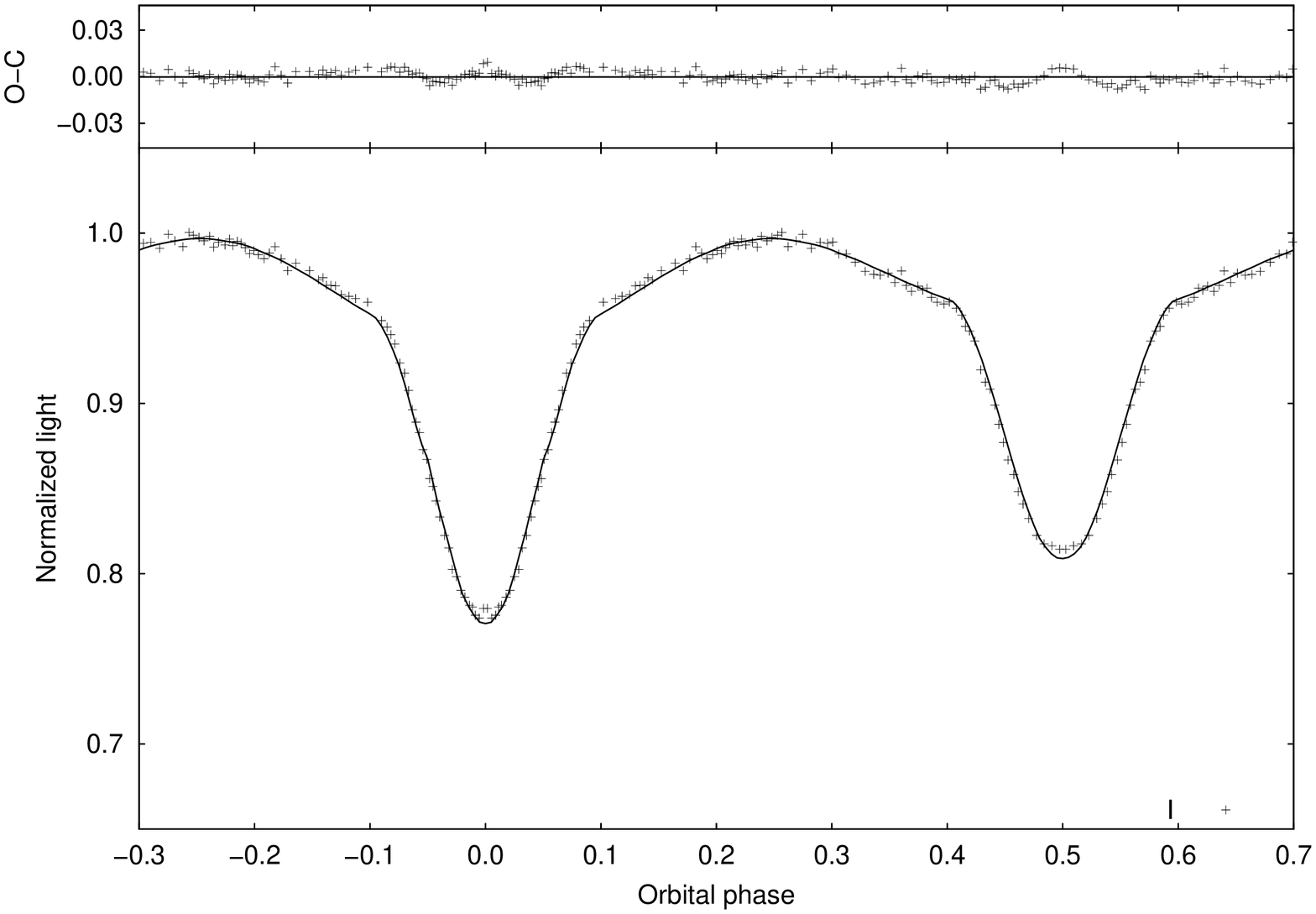}
\caption{Solution of the Wesselink light curve with WD code. In the 
top panel, the differences "measurements minus theoretical curve" are plotted.}
\label{WESSE}
\end{figure}

Naturally, as the curves differ, their solutions differ too.
The largest differences are among the solutions with $q=0.71$ and
the HHH solution, with $q=0.871$. We also solved the Chochol curves with
$q=0.871$. The adopted codes (LIGHT by HHH and WD by ourselves) are 
similar, hence the  probable reason for the different results is that
the Chochol curves constrain the system parameters only weakly. We also
found that the scatter of normal points as published by Chochol is
about twice as large as that of the $UBV$ data used in LMD, and that the
coverage of both minima is superior in the $UBV$ data. 
When solving the Wesselink curve, we used the normal points given by
Wesselink. The data and their WD fit are shown in Fig.~\ref{WESSE}.

According to Table 3 in LMD, EW$_2/$EW$_1 \approx 0.26$ for the lines
\he and 4922~\Aa. When EWs for these lines are calculated using models
e.g. by \citet{lanz}, it appears that they are practically constant in 
the given range of temperatures and are only weakly dependent on 
$\log g$. Therefore, $L_2/L_1$ in $V$ should be equal to the observed 
ratio EW$_2$/EW$_1$. 

Among the light-curve solutions, only the Wesselink curve gives such a
luminosity ratio. Although we would gladly choose the solution of a 
photoelectric curve (and are not fully satisfied with the quality of our
solution of the Wesselink curve because of the not too ideal fit of the 
minima bottoms) the spectroscopic evidence is clearly more consistent 
with the Wesselink curve, and is certainly incompatible with the HHH 
solution. The corresponding 
parameters of the system are listed in Table~\ref{ABS}. 

Since the binary is well detached, we assumed that the evolution of the
binary components was not affected by binary interaction effects and can
be well compared with evolutionary models for single stars. Using 
models by \citet{cla}, with our values of $R_1$ and $T_1$ one gets the age
5.9~My and the mass 19.5~\ms. The age is close to 5.8~My given by 
\citet[][based on $uvby$ and H$\beta$ photometry] {maly} for the cluster
NGC~1502 and the mass is explainable by the often
observed discrepancy between dynamical and evolutionary masses. In the 
case of the secondary, Claret's model gives the age 9~My and the mass 
11.2~\ms\, for $R_2$ and $T_2$. In view of the possible errors, this model
is acceptable. A larger $R_2$, which would follow from other solutions
in Table~\ref{LC}, would infer a yet older age.

The magnitude difference of the third body given by \citet{balega}
corresponds to the third body contribution of 29.4~\%, i.e. agrees well
with most of the solutions.
Summing up the absolute visual magnitudes of all three components, one
derives a total $M_{\rm v}=-5.02$. With $V_{\rm max}=6.98$ and
$A_{\rm v}=2.27$ \citep[both numbers according to][]{craw}, the distance
modulus is $m-M=9.73$, close to Crawford's value of 9.71 (Crawford 
obtained this value as the average moduli of 20 cluster members using
$uvby$ and H$\beta$ photometry analysis).

It is certainly possible to assume synchronized rotation in this 
close binary. Accepting the solution of the Wesselink curve, the primary
and secondary \vsin velocities should be 172~\ks\, and 98~\ks.
Using the He\,{\sc i} lines 4471~\Aa\, and 5876~\Aa\, in the ELODIE spectra,
we compared their FWHM with tables of \citet{abt} and a formula from
\citet{mun}, and derived \vsin velocities of 160~\ks\, for the primary,
92~\ks\, for the secondary, and 60~\ks\, for the tertiary components, each with an 
accuracy of about 6\,\%. The agreement is not perfect, but better than
for other light-curve solutions.

\begin{table}
\caption{Absolute parameters of the eclipsing system.}
\begin{minipage}[]{\textwidth}
\label{ABS}
\begin{tabular}{lcccc}
\hline\noalign{\smallskip}
Parameter            & Primary & m.e. & Secondary & m.e. \\[1mm]
\hline\noalign{\smallskip}
Semiaxis (\rs)       & 10.4    & 0.1  & 14.4      & 0.1  \\
Mass (M$_{\odot}$)   & 16.6    & 0.4  & 11.9      & 0.3  \\
Radius (R$_{\odot}$) &  9.4    & 0.2  &  5.4      & 0.2  \\
$T_{\rm eff}$ (K)    & 30\,500 & fixed& 25\,300   & 740  \\
$\log L/L_{\odot}$   &  4.837  & 0.005&  4.030    & 0.009\\
$M_{\rm bol}$        &$-7.34$  & 0.04 &$-5.33$    & 0.05 \\
BC (mag)             &$-2.93$\footnote{\citet{mar}.}  &  
    &$-2.51$\footnote{\citet{lanz}.}&\\
$M_{\rm v}$          &$-4.41$  & 0.04 &$-2.82$    & 0.05 \\
\hline\noalign{\smallskip}
\end{tabular}
\end{minipage}
\end{table}

\section{Conclusions}
We have attempted to explain the discrepancy between the $K_1$ values of \sz
obtained by HHH and ourselves by the inadequacy of the CCF method for the
solution of multiple systems. HHH also found unrealistic velocities for
the third body: absolute values of their third-body radial velocities
are larger than 40 \ks, i.e.
they disagree with the values given by LMD, \citet{g2002, g2008}, and this
paper. The CCF method produced erroneous results in the case of 
another binary in the HHH paper, IU~Aur: no third light was found,
although $l_3 \approx 30\,\%$ was deduced by various ways in this case.

In a broader context, problems with the application of the CCF method were
also highlighted by \citet{ruc}. He mentioned namely the unreliable
results achieved when the method is applied to three-body systems, and
pointed out the uncertainty in the relative luminosity determination -
both problems now being illustrated in the case of SZ~Cam. In the cases
of SZ~Cam and IU~Aur, the low quality (S/N of only 40 and 20,
respectively) and low number of spectra used by HHH might play a role.

In several papers, the CCF method has been used to study extragalactic 
binaries. It is rather likely that a third component is often present in these
binaries because of spatially unresolved field stars, and the results of the
CCF method especially for binaries with light curves of small amplitude
might be doubtful.

\begin{acknowledgement}
We are obliged to colleagues from the Ond\v rejov Observatory and their 
guests who took the SZ~Cam and ADS 2984~A spectra (Drs. M. Ceniga, P.
Hadrava, A. Kawka, D. Kor\v c\'akov\'a, L. Kotkov\'a, B. Ku\v cerov\'a, 
G. Michalska, E. Niemczura, V. Votruba, P. Zasche). PM was supported by
the Research Project MSM0021620860 of the Ministry of Education, Czech
Republic.
\end{acknowledgement}


\begin{thebibliography}{}

\bibitem[Abt et al.(2002)]{abt} Abt, H.A., Levato, H., \& Grosso, M. 2002,
       ApJ, 573, 359

\bibitem[Balega et al.(2007)]{balega} Balega, I.I., Balega, Yu.Yu., Maximov,
       A.F., et al. 2007, Astrophys. Bull., 62, 339

\bibitem[Budding(1978)]{budd} Budding, E. 1978, Ap\&SS, 36, 329

\bibitem[Chochol(1980)]{cho} Chochol, D. 1980, BAICz, 31, 321

\bibitem[Claret(2004)]{cla} Claret, A. 2004, A\&A, 424, 919

\bibitem[Conti \& Alschuler(1971)]{conti} Conti, P., \& Alschuler, W.R. 
       1971, ApJ, 170, 325

\bibitem[Crawford(1994)]{craw} Crawford, D.L. 1994, PASP, 106, 397

% \bibitem[Daflon et al.(2007)]{daf} Daflon, S., Cunha, K., de Ara\'ujo, F.X.,
%       Wolff, S., \& Przybilla, N. 2007, AJ, 134, 1570

\bibitem[ESA(1997)]{ESA} ESA 1997, The Hipparcos and Tycho Catalogues, ESA
       SP-1200, ESA, Noordwijk, the Netherlands

\bibitem[Gorda(2000)]{g2000} Gorda, S.Yu. 2000, IBVS, 4839

\bibitem[Gorda(2002)]{g2002} Gorda, S.Yu. 2002, IBVS, 5345

\bibitem[Gorda(2008)]{g2008} Gorda, S.Yu. 2008, Astrophys. Letters, 34, 1

\bibitem[Gorda et al.(2007)]{g2007} Gorda, S.Yu., Balega, Yu.Yu., Pluzhnik,
       E.A., \& Shkhagosheva, Z.U. 2007, Astrophys. Bull., 62, 352

\bibitem[Harries et al.(1998)]{HHH} Harries, T.J., Hilditch, R.W., \& Hill,
       G. 1998, MNRAS, 295, 386 (HHH)

%\bibitem{}Hohle, M.M., Eisenbeiss, T., Mugrauer, M., et al. 2009, AN,
%       330, 511

\bibitem[Kitamura \& Yamasaki(1972)]{KY} Kitamura, M., \& Yamasaki, A. 1972,
       Tokyo Astr. Bull., No. 220

\bibitem[Lanz \& Hubeny(2003)]{lanzO} Lanz, T., \& Hubeny, I. 2003, ApJS, 146, 417

\bibitem[Lanz \& Hubeny(2007)]{lanz} Lanz, T., \& Hubeny, I. 2007, ApJS, 169, 83

\bibitem[Lorenz et al.(1998)]{LMD} Lorenz, R., Mayer, P., \& Drechsel, H. 1998, 
       A\&A, 332, 909 (LMD)

% \bibitem[Lorenz et al.(2005)]{l2005} Lorenz, R., Mayer, P., \& Drechsel,
         H. 2005, MNRAS, 360, 915

\bibitem[Malysheva(1997)]{maly} Malysheva, L.K. 1997, PAZh, 23,667

\bibitem[Martins et al.(2005)]{mar} Martins, F., Schaerer, D., \& Hillier, D.J.
       2005, A\&A, 436, 1049

\bibitem[Mason et al.(1998)]{mason} Mason, D.R., Gies, D.R., William I.H., et 
       al. 1998, AJ, 115, 821

\bibitem[Massey et al.(2005)]{massey} Massey, P., Puls, J., Pauldrach, A.W.A.,
       et al. 2005, ApJ, 627, 477

\bibitem[Mayer et al.(1994)]{m94} Mayer, P., Lorenz, R., Chochol, D., \& 
       Irsmambetova, T.R. 1994, A\&A, 288, L13

%\bibitem{}Mayer, P., Wolf, M., Niarchos, P.G., et al. 2006, Ap\&SS, 304, 39

\bibitem[Michalska et al.(2007)]{mich} Michalska, G., Kub\'at, J., 
       Kor\v c\'akov\'a, D., et al. 2007, IAUS, 240, 555

%\bibitem{}Morgan, W.W., Code, A.D, \& Whitford, A.E. 1955, ApJS, 2, 41

\bibitem[Moultaka et al.(2004)]{moul} Moultaka, J., Ilovaiski, S.A., Pruguiel,
       P., \& Soubiran, C. 2004, PASP, 116, 693

\bibitem[Munari \& Tomasella(1999)]{mun} Munari, U., \& Tomasella, L. 1999,
       A\&A, 343, 806

\bibitem[Rucinski(2002)]{ruc} Rucinski, S.M. 2002, AJ, 124, 1746

% \bibitem[Simon-Diaz \& Herrero(2007)]{sim} Simon-Diaz, S., \& Herrero, A.
%       2007, A\&A, 468, 1063

\bibitem[Wesselink(1941)]{wess} Wesselink, A.J. 1941, Leiden Ann., 17, No. 3

\end{thebibliography}
\end{document}